\documentclass[a4paper]{article}
\usepackage{amsmath}
\usepackage{amssymb}
\usepackage{graphicx}
\usepackage{wrapfig}
\usepackage{enumitem}
\usepackage{float}
\usepackage{geometry}
\usepackage{quantikz}
\usepackage{tabularx}
\usepackage{lscape}
\usepackage{rotating}
\usepackage{array}
\usepackage{caption}
\usepackage{wrapfig}
\usepackage{hyperref}
\usepackage{multirow}
\usepackage{makecell}
\usepackage{rotating}
\usepackage{longtable}
\usepackage{subcaption}
\usepackage{hyperref}
\newcommand{\citep}[1]{\cite{#1}}

\date{}  
\title{Quantum Data Encoding: A Comparative Analysis of Classical-to-Quantum Mapping Techniques and Their Impact on Machine Learning Accuracy}

\begin{document}
	
	\newgeometry{
		left=3cm, 
		right=3cm, 
	}
	
\maketitle
\vspace{-1.5cm}
\author{\begin{center}
		Minati Rath\textsuperscript{*1} and Hema Date\textsuperscript{2}\\
		minati.rath.2019@iimmumbai.ac.in, hemadate@iimmumbai.ac.in\\
		\textsuperscript{1}Department of Decision Science, IIM Mumbai, India\\
		\textsuperscript{2}Department of Decision Science, IIM Mumbai, India
	\end{center}}

\begin{abstract}
This research explores the integration of quantum data embedding techniques into classical machine learning (ML) algorithms, aiming to assess the performance enhancements and computational implications across a spectrum of models. We explore various classical-to-quantum mapping methods, ranging from basis encoding, angle encoding to amplitude encoding for encoding classical data, we conducted an extensive empirical study encompassing popular ML algorithms, including Logistic Regression, K-Nearest Neighbors, Support Vector Machines, and ensemble methods like Random Forest, LightGBM, AdaBoost, and CatBoost.
Our findings reveal that quantum data embedding contributes to improved classification accuracy and F1 scores, particularly notable in models that inherently benefit from enhanced feature representation. We observed nuanced effects on running time, with low-complexity models exhibiting moderate increases and more computationally intensive models experiencing discernible changes. Notably, ensemble methods demonstrated a favorable balance between performance gains and computational overhead.

This study underscores the potential of quantum data embedding in enhancing classical ML models and emphasizes the importance of weighing performance improvements against computational costs. Future research directions may involve refining quantum encoding processes to optimize computational efficiency and exploring scalability for real-world applications. Our work contributes to the growing body of knowledge at the intersection of quantum computing and classical machine learning, offering insights for researchers and practitioners seeking to harness the advantages of quantum-inspired techniques in practical scenarios.
\end{abstract}
\section{Introduction}

The realm of computing stands at the brink of a significant transformation with the emergence of quantum technologies\cite{national2018quantum}. Quantum computing, characterized by the exploitation of quantum mechanical phenomena, such as superposition and entanglement, offers the promise of fundamentally altering the way we process and analyze information \cite{Feynman1982}\cite{Smith2023}. In this context, the translation of classical data into quantum representations has emerged as a novel and captivating avenue of exploration \cite{Johnson2022}.

Machine learning, which has already revolutionized fields ranging from image recognition to healthcare diagnostics \cite{he2019deep}\cite{rajpurkar2017chexnet}\cite{gulshan2016development}, is expected to reap substantial benefits from the integration of classical and quantum computing. The potential to enhance machine learning models through the utilization of quantum data representations raises a host of intriguing questions and possibilities. How can classical data, which has long served as the backbone of data-driven decision-making, be seamlessly and effectively translated into the quantum realm. What quantum data encoding techniques yield the most promising results for classical machine learning models. \cite{Brown2021}

This research embarks on an empirical journey to address these questions. The primary objective is to investigate the efficacy of classical data translation into quantum states using a variety of encoding techniques. These techniques encompass a spectrum of classical-to-quantum data mapping methods, including basis encoding, angle encoding and amplitude encoding. The unique impact of each technique on classical machine learning performance when applied to both classical and quantum datasets are rigorously examined.

Of particular significance is our approach, which centers on maintaining data consistency. By utilizing the same dataset across all experiments, we ensure that the evaluation of classical and quantum data representations remains as objective and informative as possible. This meticulous method allows us to isolate the effects of the encoding techniques themselves, unraveling the intricate relationship between quantum data representations and machine learning performance.

In our pursuit, we aim to provide comprehensive insights into the practical implications of quantum data encoding for classical machine learning. This research, through systematic experimentation and comparative analysis, offers guidance to researchers and practitioners seeking to harness the potential of quantum technologies to augment classical data analysis. Furthermore, it contributes to the broader conversation surrounding the integration of quantum and classical computing, providing nuanced perspectives on how quantum data representations can enhance classical machine learning models.

This study serves as a timely exploration into the opportunities and challenges presented by quantum data encoding, shedding light on the path forward as we navigate the intersection of classical and quantum computing.

\section{Literature Review}

The intersection of classical machine learning and quantum computing has garnered significant interest in recent years, as it holds the potential to revolutionize the landscape of data-driven decision-making. This literature review provides an overview of the key developments in the fields of quantum data encoding and classical machine learning, highlighting the evolution of this exciting fusion.

Quantum data encoding techniques form the foundation of this research, offering a gateway to translating classical data into quantum representations. One of the earliest pioneering works in this field, Feynman's proposition of quantum computing in 1982, set the stage for quantum data encoding. Feynman highlighted the potential of quantum systems to simulate physical systems \cite{Feynman1982} efficiently, leading to the conceptualization of quantum algorithms for various applications. These ideas laid the groundwork for the exploration of quantum data encoding techniques in subsequent research.

A seminal work by John Preskill introduced the concept of NISQ (Noisy Intermediate-Scale Quantum) devices, marking a significant milestone in the development of practical quantum computing. This paper emphasized the potential of quantum computers, even in their early stages of development, to outperform classical computers in certain applications\cite{national2018quantum}.

The intigration of classical machine learning with quantum computing is a compelling frontier. Michael Brown's 2021 paper, "Machine Learning Models for Quantum Data," explores the potential of enhancing classical machine learning models using quantum data representations. Brown's work emphasizes the need to harness the unique capabilities of quantum systems to advance classical data analysis \cite{Brown2021}.

While there is a growing body of work on quantum data encoding and classical machine learning, there remains a notable gap in research that systematically investigates the effectiveness of quantum data encoding techniques in enhancing classical machine learning models. This paper addresses this gap by undertaking an empirical study that comprehensively explores the impact of different encoding methods on classical machine learning performance.

In summary, the integration of quantum and classical computing offers a promising avenue for data analysis. However, comprehensive empirical studies that investigate the practical implications of quantum data encoding for classical machine learning are relatively scarce. This research aims to contribute to this growing discourse by conducting rigorous experiments and comparative analysis to shed light on the opportunities and challenges presented by quantum data encoding.
\section{Quantum Computing}
Quantum computing exhibits a set of remarkable properties that differentiate it from classical computing. Chief among these is superposition, which allows quantum bits, or qubits, to exist in multiple states simultaneously \cite{nielsen2000quantum}, enabling parallel computations \cite{aaronson2011quantum}. Entanglement is another key property, where the state of one qubit is intrinsically linked to another, even when separated by vast distances, offering the potential for distributed quantum computing. Quantum interference leverages the unique probability amplitudes of quantum states to enhance computational efficiency. The no-cloning theorem ensures the security of quantum cryptography by prohibiting the perfect cloning of an unknown quantum state. Quantum tunneling facilitates problem-solving by overcoming classical barriers \cite{venegas2004quantum}, while quantum parallelism evaluates multiple possibilities concurrently, potentially providing faster solutions for specific problems. Quantum algorithms like Shor's and Grover's promise exponential speedup\cite{shor1997}\cite{grover1996}, and error correction codes are crucial for maintaining computation reliability\cite{gottesman1997quantum}. Although universal quantum computation capabilities for all problems are still under exploration, quantum computing's distinct properties hold significant promise for specific applications.
\section{Quantum Data}
Quantum data refers to information or data that is stored and processed using quantum bits, or qubits, and adheres to the principles of quantum mechanics. Quantum data can be represented in various forms, such as quantum states, quantum registers, or quantum circuits. Unlike classical data, which is typically represented in binary format (0s and 1s), quantum data can exist in multiple states simultaneously due to the property of superposition. This allows quantum data to encode and process a broader range of information efficiently.

Furthermore, quantum data can exhibit entanglement, meaning the state of one qubit is correlated with the state of another, even when separated by considerable distances. This property enables the creation of entangled quantum data for applications like quantum cryptography and quantum teleportation.

Quantum data encoding techniques play a crucial role in translating classical data into quantum states. These techniques encompass various methods, including basis encoding, angle encoding, amplitude encoding, quantum autoencoders \cite{9256033}, and quantum convolutional neural networks \cite{Cong2019}, each with its advantages and applications. Quantum data holds promise in fields like quantum computing, quantum cryptography \cite{Yin2020}, and quantum machine learning, where its unique properties can be harnessed to perform computations and secure communications more efficiently than classical data.

Quantum data, like classical data, comes in various types and forms, each with its own unique characteristics and applications. Here are some common types of quantum data: \\
Quantum States - Quantum data often refers to the state of a quantum system, which can be described using quantum states. These states can represent qubits, qudits (higher-dimensional quantum systems), or even more complex quantum systems. Examples include the quantum state $|0\rangle$, $|1\rangle$, superposition states (e.g.,$|0\rangle + |1\rangle$), and entangled states. \\
Quantum Registers - Quantum data can be organized into quantum registers, which are collections of qubits or qudits. Quantum registers play a fundamental role in quantum algorithms and computations, and the data within these registers can be manipulated collectively.\\
Quantum Gates - Quantum gates manipulats quantum bits or qubits to perform computations. Among the elementary gates are the Pauli gates, including the X gate that flips qubit states. The Hadamard gate induces superposition, placing qubits in a probabilistic mix of |0⟩ and |1⟩. Phase gates, like the S gate, introduce phase shifts, rotating qubit states. The CNOT gate entangles two qubits, while the SWAP gate exchanges their states. Toffoli gates act as controlled-controlled NOT gates, flipping a target qubit based on the states of two control qubits. The U gate represents a general single-qubit rotation, and Rz, Rx, and Ry gates perform rotations around the Z, X, and Y axes, respectively. Quantum circuits leverage these gates, each represented by a unitary matrix, to execute specific algorithms efficiently. Shor's and Grover's algorithms, among others, showcase the power of these gates in solving problems beyond classical capabilities.\\
Quantum Circuits - Quantum data can be processed and transformed through quantum circuits. These circuits represent sequences of quantum gates that manipulate the quantum state. Quantum circuits are central to quantum algorithms and quantum computations.\\

\setlength{\intextsep}{1pt} 
\begin{wrapfigure}{r}{0.4\textwidth}
	\centering
	\includegraphics[width=1\linewidth]{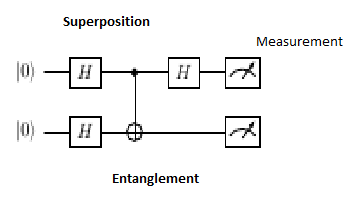}
	\caption{Quantum Circuit}
	\label{fig:QuantumCircuit}
\end{wrapfigure}

Quantum Measurements - The outcomes of quantum measurements constitute quantum data. Quantum measurements reveal information about the quantum state, and the results are often probabilistic due to the fundamental properties of quantum mechanics.\\

Quantum Entanglement - Quantum data can include entangled states. Entanglement is a unique form of quantum correlation where the state of one particle is intrinsically linked to the state of another, regardless of the distance between them. Measuring one entangled particle instantaneously determines the state of the other.\\

Quantum Keys - In the context of quantum cryptography, quantum data refers to quantum keys, which are cryptographic keys generated using quantum properties like entanglement and the no-cloning theorem. These keys are used for secure communication \cite{lutkenhaus2000quantum} \cite{ablayev2017quantum}.  Quantum keys are typically generated using Quantum Key Distribution protocols, such as the well-known BBM92 \cite{BB84} (named after its inventors, Charles Bennett, Gilles Brassard, and Artur Ekert in 1992) or more recent protocols like E91 \cite{Ekert91}. QKD leverages the principles of quantum mechanics to secure the key exchange process against eavesdropping.\\

Quantum Images and Quantum Video - Quantum data can be extended to quantum image and video formats, allowing for the storage and transmission of visual information using quantum properties.

\subsection{Quantum Data Encoding Schemes}

\subsubsection{Basis Encoding (or Computational Basis Encoding)} 
This is the most straightforward form of quantum encoding. In this method, classical bits are directly mapped to quantum bits (qubits) using the computational basis states \( |0\rangle \) and \( |1\rangle \). The information is stored in the amplitudes of these states. A classical number 101 can be represented as the quantum state \( |101\rangle \):
\[|101\rangle = |1\rangle \otimes |0\rangle \otimes |1\rangle.\]
In terms of amplitudes:
\[|101\rangle = \alpha |1\rangle \otimes \beta |0\rangle \otimes \gamma |1\rangle,\]
where \( \alpha \), \( \beta \), \( \gamma \) are the probability amplitudes associated with each qubit state.  The tensor product symbol $\otimes $ indicates that each qubit is considered independently.\\
The basis encoding of the word "hello" is given by the quantum states:
$|h\rangle = |1101000\rangle, \quad |e\rangle = |1100101\rangle, \quad |l_1\rangle = |1101100\rangle, \quad |l_2\rangle = |1101100\rangle, \quad |o\rangle = |1101111\rangle $
To perform basis encoding of the word "hello," each ASCII character is converted into its binary representation and then encode each bit using quantum states.ASCII for 'h': 104 (binary: 1101000), ASCII for 'e': 101 (binary: 1100101), ASCII for 'l': 108 (binary: 1101100), ASCII for 'l': 108 (binary: 1101100), ASCII for 'o': 111 (binary: 1101111).  This encoding scheme is suitable for discrete data represented as binary or integer values.

\subsubsection{Superposition Encoding} Quantum superposition allows qubits to exist in multiple states simultaneously. Information can be encoded by putting qubits in a superposition of classical states, thereby encoding multiple possibilities at once.  Superposition encoding is represented a quantum state as a linear combination of basis states.
\begin{align*}
	|101\rangle &= \frac{1}{\sqrt{3}}|100\rangle + \frac{1}{\sqrt{3}}|010\rangle + \frac{1}{\sqrt{3}}|001\rangle \\
	|78\rangle &= \frac{1}{\sqrt{2}}|1001110\rangle + \frac{1}{\sqrt{2}}|0100111\rangle \\
	|hello\rangle &= \frac{1}{\sqrt{5}}|1101000\rangle + \frac{1}{\sqrt{5}}|0010011\rangle + \frac{1}{\sqrt{5}}|1011011\rangle + \frac{1}{\sqrt{5}}|1100110\rangle + \frac{1}{\sqrt{5}}|1101111\rangle
\end{align*}

\subsubsection{Angle Encoding} Classical information is encoded in the relative phase between different quantum states. A phase shift is usually represented as a rotation in the complex plane.  This encoding takes advantage of the fact that the probability of measuring a particular state is determined by its phase.  Quantum states are represented by complex numbers, and their amplitudes correspond to the probabilities of measuring different outcomes. The phase of these amplitudes, expressed as angles, plays a crucial role in quantum algorithms.  In parameterized quantum circuits (PQCs), the parameters of a quantum circuit are adjusted to execute specific computations. Angle encoding serves as a method for integrating classical information into quantum states, enabling quantum computers to effectively handle and manipulate classical data.  Quantum gates use rotation operations around different axes (X, Y, or Z) for changing quantum state . For example, the rotation around the X-axis, denoted as \(R_x(\theta)\), is represented as:
\[ R_x(\theta) = e^{-i\theta X/2} =
\begin{bmatrix}
	\cos(\theta/2) & -i\sin(\theta/2) \\
	-i\sin(\theta/2) & \cos(\theta/2)
\end{bmatrix}
\]

Similarly, rotation around the Y-axis denoted as  \(R_y(\theta)\) is
\[ R_y(\theta) = e^{-i\theta Y/2} =
\begin{bmatrix}
	\cos(\theta/2) & -\sin(\theta/2) \\
	\sin(\theta/2) & \cos(\theta/2)
\end{bmatrix} \]
 and rotation around the Z-axis denoted as  $R_z(\theta)$
 \[ R_z(\theta) = e^{-i\theta Z/2} =
 \begin{bmatrix}
 	e^{-i\theta/2} & 0 \\
 	0 & e^{i\theta/2}
 \end{bmatrix} \]

It's important to note that the encoding scheme might involve specific quantum gates like the RX gate. Using the RX gate for encoding involves choosing appropriate angles for the rotation.

The angle embedding in quantum computing is often represented mathematically as follows: Let's consider a single qubit state \(\ket{\psi}\) and an angle \(\theta\). The angle embedding operation can be expressed as:
\[ \ket{\psi(\theta)} = R_y(\theta) \ket{0} + e^{i\phi} R_y(\theta) \ket{1} \]
Here, \(R_y(\theta)\) is the quantum gate corresponding to a rotation around the y-axis by an angle \(\theta\), and \(\ket{0}\) and \(\ket{1}\) are the basis states of the qubit. The phase factor \(e^{i\phi}\) is optional and depends on the specific encoding scheme.

In more general terms, for a single-qubit rotation around an arbitrary axis defined by the unit vector \(\hat{n} = (n_x, n_y, n_z)\), the angle embedding can be written as:

\[
\begin{aligned}
	\ket{\psi(\theta)} = & \cos\left(\frac{\theta}{2}\right) \ket{0} + e^{i\phi}\sin\left(\frac{\theta}{2}\right) \ket{1} \\
	& + \cos\left(\frac{\theta}{2}\right) \ket{0} - e^{i\phi}\sin\left(\frac{\theta}{2}\right) \ket{1}
\end{aligned}
\]

Here, \(\phi\) is the phase factor, and \(X\), \(Y\), and \(I\) are Pauli matrices and the identity matrix, respectively.

This general formula allows  to perform an angle embedding operation using a rotation around any axis in the Bloch sphere. 

For instance, angle encoding the binary value \(101\) could involve specific angle values corresponding to each bit.  For the binary value \(101\), associate angles \(\theta_1\), \(\theta_2\), and \(\theta_3\) with bits \(1\), \(0\), and \(1\) respectively. The quantum state resulting from the angle encoding is:
\[ \text{Angle Encoding}(\text{binary }101) = R_x(\theta_1) R_x(\theta_2) R_x(\theta_3) |0\rangle
\]

 The binary value 101 can be encoded into a multi-qubit state using angle encoding with  $R_x$ gates, rotations are applied to each qubit based on the binary values.
\begin{align*}
	|\psi\rangle &= R_{x_2}(\theta_2) \cdot R_{x_1}(\theta_1) \cdot R_{x_0}(\theta_0) |0\rangle \\
	&= e^{-i\theta_2\sigma_{x_2}/2} \otimes e^{-i\theta_1\sigma_{x_1}/2} \otimes e^{-i\theta_0\sigma_{x_0}/2} |0\rangle,
\end{align*}
 To encode the binary value  101:
\begin{align*}
	\theta_0 &= 0 \quad \text{(no rotation for the first qubit)}, \\
	\theta_1 &= \pi \quad \text{(rotate the second qubit by 180 degrees)}, \\
	\theta_2 &= 0 \quad \text{(no rotation for the third qubit)}.
\end{align*}
So, rotation will be $|\psi\rangle = R_{x_2}(0) \cdot R_{x_1}(\pi) \cdot R_{x_0}(0) |0\rangle\ $

To angle encode integer value 78 using $R_x$ gate, 
\begin{align*}
	R_x(78) &= \begin{bmatrix} \cos\left(\frac{78}{2}\right) & -i\sin\left(\frac{78}{2}\right) \\ -i\sin\left(\frac{78}{2}\right) & \cos\left(\frac{78}{2}\right) \end{bmatrix} \\
	R_x(78^\circ) &\approx \begin{bmatrix} 0.766 & -0.642i \\ -0.642i & 0.766 \end{bmatrix}\\
	 &\approx [0.766  -0.642i]
\end{align*}

\subsubsection{Amplitude Encoding} In this scheme, classical Information is encoded in the amplitudes of quantum state. By adjusting the probability amplitudes of different states, information can be represented in a quantum superposition. In contrast to basis encoding, where information is encoded in the probability distribution of measurement outcomes, amplitude encoding utilizes the amplitudes of the quantum state itself.  In amplitude encoding, the probability of measuring a particular outcome is determined by the squared magnitude of the corresponding amplitude. The amplitudes can be manipulated to encode classical information in a quantum superposition. A quantum state \(|\psi\rangle = \alpha|0\rangle + \beta|1\rangle\) with $\alpha$ and $\beta$ are complex numbers representing the amplitudes, classical information are encoded by adjusting these amplitudes. $\alpha$ corresponds to the classical bit value 0 and $\beta$ corresponds to the classical bit value 1, then the state $\psi$ encodes classical information \cite{nielsen2000quantum}.

To encode real values X=[1.2,2.7,1.1,0.5], two qubits are required.  $|\psi(x)\rangle = \sum\limits_{i=1}^{n} x_i |i\rangle$, where $x_i$ is the $i^{th}$ term of x.  Each term is square normalised $\frac{1}{\sqrt{10.19}}$. 

The state will be $|\psi\rangle$ = $\frac{1.2}{\sqrt{10.19}}|00\rangle$ +  $\frac{2.7}{\sqrt{10.19}}|01\rangle$ + $\frac{1.1}{\sqrt{10.19}}|10\rangle$ + $\frac{0.5}{\sqrt{10.19}}|11\rangle$.

\section{Experimental Results}
We experimented with a few machine learning algorithms using PennyLane open source software library to tackle a customer churn classification problem within the telecommunications domain. 
\\The dataset comprises 20 features, including a binary target column indicating churn, and encompasses a total of 7,043 distinct customers. This dataset is available on the Kaggle website.
\begin{wrapfigure}{R}{0.6\textwidth}
	\begin{center}
		\includegraphics[width=0.5\linewidth]{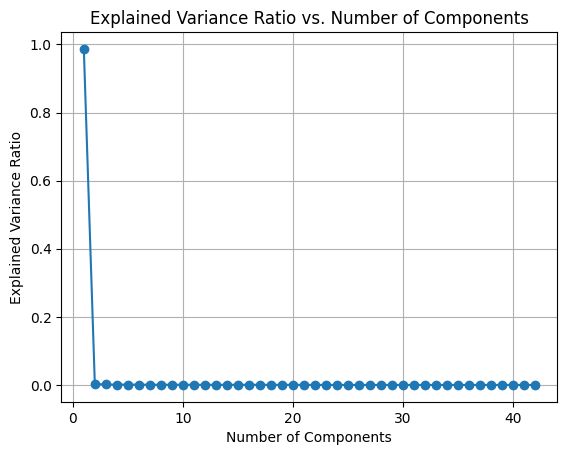}
		\caption{Explained Variance ration with Elbow point}
		\label{fig:ExplainedVarianceElbow}
	\end{center}
	\begin{quote}
		\centering
		Elbow Point Index: 23\\
		Explained Variance Ratio at Elbow Point: 8.126234391114803e-33 \\
		Cumulative Explained Variance at Elbow Point: 1.0000000000000002 \\
	\end{quote}
\end{wrapfigure} 

Among the features, the following are categorical: gender, SeniorCitizen, Partner, Dependents, PhoneService, MultipleLines, InternetService, OnlineSecurity, OnlineBackup, DeviceProtection, TechSupport, StreamingTV, StreamingMovies, Contract, PaperlessBilling, PaymentMethod, and Churn. On the other hand, tenure, MonthlyCharges, and TotalCharges are numeric features. A notable correlation coefficient of 0.83 between Tenure and TotalCharges led to the exclusion of TotalCharges from our analysis. Fortunately, no outliers or missing values were found in the dataset.

Concerning the PhoneService feature, multicollinearity variance inflation factor (VIF) scores above 12 prompted us to eliminate it from further processing. After reevaluating the VIF, MonthlyCharges, with a score close to 6, was also dropped. Additionally, the customerID feature was removed as it does not contribute to the model.

Out of the 16 features, all except 'tenure' underwent one-hot encoding, resulting in a total of 42 columns. The classes in the churn label exhibit significant imbalance, with 1,869 instances belonging to the minority 'Yes' class and 5,174 instances belonging to the majority 'No' class. Due to the limitations of the simulator in handling large volumes of data, we opted to retain the minority classes and performed undersampling on the majority classes, resulting in a total of 3,738 records.

We attempted to determine the optimal number of components where the explained variance ratio diminishes significantly. By plotting the number of components against the explained variance ratio, we identified an elbow point at 23. Figure 2 illustrates the relationship between explained variance and the number of components.  
\vspace{10pt} 

We experimented with popular ML algorithms, including Logistic Regression, K-Nearest Neighbors, Support Vector Machines, and ensemble methods like Random Forest, LightGBM, AdaBoost, and CatBoost. These machine learning algorithms encompass a diverse set of techniques. Logistic Regression, a supervised learning method, excels in binary classification by modeling class probabilities through the logistic function\cite{hosmer2013logistic}. K-Nearest Neighbors (KNN), functioning in both supervised and unsupervised contexts, is versatile for classification and regression tasks, relying on the proximity of data points\cite{yianilos1993fast}. Support Vector Machines (SVM) are powerful for classification and regression, creating decision boundaries by maximizing the margin between classes\cite{burges1998tutorial} \cite{schoelkopf2002kernel}. Ensemble methods, such as Random Forest, LightGBM, AdaBoost, and CatBoost, leverage the strength of multiple models for enhanced performance. Random Forest constructs a multitude of decision trees, combining their outputs for robust predictions\cite{breiman2001random}. Gradient Boosting Trees (GBT) builds a series of decision trees sequentially, where each tree corrects the errors of the previous one. \cite{chen2016xgboost} LightGBM optimizes gradient boosting \cite{ke2017lightgbm}, AdaBoost focuses on weak learners\cite{schapire1998boosting}, and CatBoost excels in handling categorical features\cite{dorogush2018catboost}. Collectively, these algorithms cater to various machine learning scenarios, showcasing flexibility and effectiveness in different applications.

In classification tasks, we used various evaluation metrics to assess the performance of machine learning models. The accuracy, defined as the ratio of correct predictions to the total number of predictions, provides an overall measure of classification correctness. Precision, or the positive predictive value, is the ratio of true positives to the sum of true positives and false positives, indicating the accuracy of positive predictions. Sensitivity, also known as recall or the true positive rate, measures the model's ability to capture all positive instances by dividing true positives by the sum of true positives and false negatives. The F1 score, a harmonic mean of precision and recall, balances the trade-off between these two metrics. Additionally, the Area Under the Receiver Operating Characteristic curve (ROC AUC) assesses the model's ability to discriminate between positive and negative instances across different probability thresholds. A higher ROC AUC signifies superior model performance, particularly useful when evaluating the discriminatory power of the model \cite{alpaydin2014introduction}.The Cohen's Kappa coefficient is a statistic that measures the level of agreement between two raters beyond that which would be expected by chance\cite{cohen1960coefficient}.

The table 1 below shows the performance of classical data and quantum data embedding with an 80:20 train-test split ratio.  

\vspace{2pt} 

\begin{landscape}
	\begin{longtable}{|p{1.5cm}|p{1.0cm}|p{4.7cm}|p{1.5cm}|p{1.5cm}|p{1.7cm}|p{1.0cm}|p{1.0cm}|p{1.0cm}|p{1.3cm}|p{1.5cm}|}
		\caption{Classical data and Quantum Data Embedding Performance Comparision } \label{tab:Results} \\
		\hline
		\textbf{Classifier} & \textbf{PCA} & \textbf{Data Encoding Type} & \textbf{Accuracy} & \textbf{Precision} & \textbf{Sensitivity} & \textbf{Recall} & \textbf{F1 Score} & \textbf{ROC AUC} & \textbf{Cohen's Kappa} & \textbf{Running Time} \\
		\hline
		\endfirsthead
		
		\multicolumn{11}{c}%
		{{\tablename\ \thetable{} -- continued from previous page}} \\
		\hline
		\textbf{Classifier} & \textbf{PCA} & \textbf{Data Encoding Type} & \textbf{Accuracy} & \textbf{Precision} & \textbf{Sensitivity} & \textbf{Recall} & \textbf{F1 Score} & \textbf{ROC AUC} & \textbf{Cohen's Kappa} & \textbf{Running Time} \\
		\hline
		\endhead
		
		\hline \multicolumn{11}{|r|}{{Continued on next page}} \\ \hline
		\endfoot
		
		\hline
		\endlastfoot
		
\multirow{12}{*}{\makecell[{{p{1.5cm}}}]{Logistic \\ Regression}}  & \multirow{4}{*}{\makecell[{{p{1.0cm}}}]{2}} & Classical & 72.9947 & 0.7241 & 0.7546 & 0.7546 & 0.7390 & 0.7296 & 0.4595 & 0.0052 \\
& & Quantum Basis Encoding & 68.0481 & 0.6651 & 0.7441 & 0.7441 & 0.7024 & 0.6796 & 0.3598 & 0.0059 \\
& & Quantum Angle Encoding & 68.0481 & 0.6651 & 0.7441 & 0.7441 & 0.7024 & 0.6796 & 0.3598 & 0.0052 \\
& & Quantum Amplitude Encoding & 68.0481 & 0.6651 & 0.7441 & 0.7441 & 0.7024 & 0.6796 & 0.3598 & 0.0052 \\
\cline{2-11}
 & \multirow{4}{*}{\makecell[{{p{1.0cm}}}]{15}} & Classical & 75.9358 & 0.7519 & 0.7836 & 0.7836 & 0.7674 & 0.7590 & 0.5184 & 0.0130 \\
& & Quantum Basis Encoding & 66.8449 & 0.6513 & 0.7441 & 0.7441 & 0.6946 & 0.6674 & 0.3355 & 0.0084 \\
& & Quantum Angle Encoding & 66.8449 & 0.6513 & 0.7441 & 0.7441 & 0.6946 & 0.6674 & 0.3355 & 0.0051 \\
& & Quantum Amplitude Encoding & 66.8449 & 0.6513 & 0.7441 & 0.7441 & 0.6946 & 0.6674 & 0.3355 & 0.0051 \\
\cline{2-11}
 & \multirow{4}{*}{\makecell[{{p{1.0cm}}}]{23}} & Classical	& 79.0106 & 0.7921 & 0.7941 &	0.7941 & 0.793 & 0.7900 & 0.5801 & 0.0095\\
& & Quantum Basis Encoding & 66.8449 & 0.6512 & 0.7440 & 0.7440 & 0.6945 & 0.6674 &	0.3354	&0.0061\\
& & Quantum Angle Encoding & 66.8449 & 0.6512 & 0.7440 & 0.7440 & 0.6945 & 0.6674 &	0.3354	&0.0029\\
& & Quantum Amplitude Encoding & 66.8449 & 0.6512 & 0.7440 & 0.74406 &	0.6945 & 0.6674 & 0.3354	&0.0029\\
\hline
\multirow{12}{*}{\makecell[{{p{1.5cm}}}]{KNN}}  & \multirow{4}{*}{\makecell[{{p{1.0cm}}}]{2}} & Classical & 72.3262 & 0.7172 & 0.7493 & 0.7493 & 0.7329 & 0.7229 & 0.4461 & 0.0437 \\
& & Quantum Basis Encoding & 50.6684 & 0.5067 & 1.0000 & 1.0000 & 0.6726 & 0.5000 & 0.0000 & 0.0520 \\
& & Quantum Angle Encoding & 50.6684 & 0.5067 & 1.0000 & 1.0000 & 0.6726 & 0.5000 & 0.0000 & 0.0859 \\
& & Quantum Amplitude Encoding & 50.6684 & 0.5067 & 1.0000 & 1.0000 & 0.6726 & 0.5000 & 0.0000 & 0.0859 \\
\cline{2-11}
& \multirow{4}{*}{\makecell[{{p{1.0cm}}}]{15}} & Classical & 69.9198 & 0.6766 & 0.7784 & 0.7784 & 0.7239 & 0.6981 & 0.3971 & 0.0995 \\
& & Quantum Basis Encoding & 66.8449 & 0.6513 & 0.7441 & 0.7441 & 0.6946 & 0.6674 & 0.3355 & 0.1338 \\
& & Quantum Angle Encoding & 66.8449 & 0.6513 & 0.7441 & 0.7441 & 0.6946 & 0.6674 & 0.3355 & 0.0493 \\
& & Quantum Amplitude Encoding & 66.8449 & 0.6513 & 0.7441 & 0.7441 & 0.6946 & 0.6674 & 0.3355 & 0.0493 \\
\cline{2-11}
& \multirow{4}{*}{\makecell[{{p{1.0cm}}}]{23}} & Classical & 73.7968 & 0.7194 & 0.7916 & 0.7916 & 0.7538 & 0.7372 & 0.4751 & 0.0848 \\
& & Quantum Basis Encoding & 50.6684 & 0.5067 & 1.0000 & 1.0000 & 0.6726 & 0.5000 & 0.0000 & 0.0884 \\
& & Quantum Angle Encoding & 50.6684 & 0.5067 & 1.0000 & 1.0000 & 0.6726 & 0.5000 & 0.0000 & 0.0448 \\
& & Quantum Amplitude Encoding & 50.6684 & 0.5067 & 1.0000 & 1.0000 & 0.6726 & 0.5000 & 0.0000 & 0.0448 \\
\hline 
\multirow{4}{*}{\makecell[{{p{1.5cm}}}]{SVM Linear}}  & \multirow{4}{*}{\makecell[{{p{1.0cm}}}]{2}} & Classical & 73.1283 & 0.7247 & 0.7573 & 0.7573 & 0.7406 & 0.7309 & 0.4621 & 0.2453 \\
& & Quantum Basis Encoding & 68.0481 & 0.6651 & 0.7441 & 0.7441 & 0.7024 & 0.6796 & 0.3598 & 0.3365 \\
& & Quantum Angle Encoding & 68.0481 & 0.6651 & 0.7441 & 0.7441 & 0.7024 & 0.6796 & 0.3598 & 0.3376 \\
& & Quantum Amplitude Encoding & 68.0481 & 0.6651 & 0.7441 & 0.7441 & 0.7024 & 0.6796 & 0.3598 & 0.3376 \\
\cline{2-11}
\pagebreak
\multirow{8}{*}{\makecell[{{p{1.5cm}}}]{SVM Linear}}& \multirow{4}{*}{\makecell[{{p{1.0cm}}}]{15}} & Classical & 75.0000 & 0.7388 & 0.7836 & 0.7836 & 0.7606 & 0.7495 & 0.4995 & 0.5831 \\
& & Quantum Basis Encoding & 66.8449 & 0.6513 & 0.7441 & 0.7441 & 0.6946 & 0.6674 & 0.3355 & 0.5701 \\
& & Quantum Angle Encoding & 66.8449 & 0.6513 & 0.7441 & 0.7441 & 0.6946 & 0.6674 & 0.3355 & 0.2929 \\
& & Quantum Amplitude Encoding & 66.8449 & 0.6513 & 0.7441 & 0.7441 & 0.6946 & 0.6674 & 0.3355 & 0.2929 \\
\cline{2-11}
& \multirow{4}{*}{\makecell[{{p{1.0cm}}}]{23}} & Classical & 77.5401 & 0.7482 & 0.8391 & 0.8391 & 0.7910 & 0.7745 & 0.5500 & 0.6471 \\
& & Quantum Basis Encoding & 66.8449 & 0.6513 & 0.7441 & 0.7441 & 0.6946 & 0.6674 & 0.3355 & 0.5428 \\
& & Quantum Angle Encoding & 66.8449 & 0.6513 & 0.7441 & 0.7441 & 0.6946 & 0.6674 & 0.3355 & 0.1568 \\
& & Quantum Amplitude Encoding & 66.8449 & 0.6513 & 0.7441 & 0.7441 & 0.6946 & 0.6674 & 0.3355 & 0.1568 \\
\hline
\multirow{12}{*}{\makecell[{{p{1.5cm}}}]{SVM Poly}}  & \multirow{4}{*}{\makecell[{{p{1.0cm}}}]{2}} & Classical & 71.7914 & 0.6714 & 0.8681 & 0.8681 & 0.7572 & 0.7159 & 0.4335 & 0.3055 \\
& & Quantum Basis Encoding & 68.0481 & 0.6651 & 0.7441 & 0.7441 & 0.7024 & 0.6796 & 0.3598 & 0.4763 \\
& & Quantum Angle Encoding & 68.0481 & 0.6651 & 0.7441 & 0.7441 & 0.7024 & 0.6796 & 0.3598 & 0.3603 \\
& & Quantum Amplitude Encoding & 68.0481 & 0.6651 & 0.7441 & 0.7441 & 0.7024 & 0.6796 & 0.3598 & 0.3603 \\
\cline{2-11}
& \multirow{4}{*}{\makecell[{{p{1.0cm}}}]{15}} & Classical & 72.1925 & 0.7031 & 0.7810 & 0.7810 & 0.7400 & 0.7211 & 0.4429 & 0.3926 \\
& & Quantum Basis Encoding & 66.8449 & 0.6513 & 0.7441 & 0.7441 & 0.6946 & 0.6674 & 0.3355 & 0.4803 \\
& & Quantum Angle Encoding & 66.8449 & 0.6513 & 0.7441 & 0.7441 & 0.6946 & 0.6674 & 0.3355 & 0.2430 \\
& & Quantum Amplitude Encoding & 66.8449 & 0.6513 & 0.7441 & 0.7441 & 0.6946 & 0.6674 & 0.3355 & 0.2430 \\
\cline{2-11}
& \multirow{4}{*}{\makecell[{{p{1.0cm}}}]{23}} & Classical & 74.0642 & 0.7273 & 0.7810 & 0.7810 & 0.7532 & 0.7401 & 0.4807 & 0.4581 \\
& & Quantum Basis Encoding & 66.8449 & 0.6513 & 0.7441 & 0.7441 & 0.6946 & 0.6674 & 0.3355 & 0.7717 \\
& & Quantum Angle Encoding & 66.8449 & 0.6513 & 0.7441 & 0.7441 & 0.6946 & 0.6674 & 0.3355 & 0.1801 \\
& & Quantum Amplitude Encoding & 66.8449 & 0.6513 & 0.7441 & 0.7441 & 0.6946 & 0.6674 & 0.3355 & 0.1801 \\
\hline
\multirow{12}{*}{\makecell[{{p{1.5cm}}}]{SVM RBF}}  & \multirow{4}{*}{\makecell[{{p{1.0cm}}}]{2}} & Classical & 75.4011 & 0.7468 & 0.7784 & 0.7784 & 0.7623 & 0.7537 & 0.5077 & 0.3022 \\
& & Quantum Basis Encoding & 68.0481 & 0.6651 & 0.7441 & 0.7441 & 0.7024 & 0.6796 & 0.3598 & 1.2611 \\
& & Quantum Angle Encoding & 68.0481 & 0.6651 & 0.7441 & 0.7441 & 0.7024 & 0.6796 & 0.3598 & 0.5284 \\
& & Quantum Amplitude Encoding & 68.0481 & 0.6651 & 0.7441 & 0.7441 & 0.7024 & 0.6796 & 0.3598 & 0.5284 \\
\cline{2-11}
& \multirow{4}{*}{\makecell[{{p{1.0cm}}}]{15}} & Classical & 73.6631 & 0.7241 & 0.7757 & 0.7757 & 0.7490 & 0.7361 & 0.4727 & 0.4317 \\
& & Quantum Basis Encoding & 66.8449 & 0.6513 & 0.7441 & 0.7441 & 0.6946 & 0.6674 & 0.3355 & 0.9169 \\
& & Quantum Angle Encoding & 66.8449 & 0.6513 & 0.7441 & 0.7441 & 0.6946 & 0.6674 & 0.3355 & 0.2927 \\
& & Quantum Amplitude Encoding & 66.8449 & 0.6513 & 0.7441 & 0.7441 & 0.6946 & 0.6674 & 0.3355 & 0.2927 \\
\cline{2-11}
& \multirow{4}{*}{\makecell[{{p{1.0cm}}}]{23}} & Classical & 77.2727 & 0.7700 & 0.7863 & 0.7863 & 0.7781 & 0.7725 & 0.5452 & 0.5184 \\
& & Quantum Basis Encoding & 66.8449 & 0.6513 & 0.7441 & 0.7441 & 0.6946 & 0.6674 & 0.3355 & 0.9918 \\
& & Quantum Angle Encoding & 66.8449 & 0.6513 & 0.7441 & 0.7441 & 0.6946 & 0.6674 & 0.3355 & 0.2802 \\
& & Quantum Amplitude Encoding & 66.8449 & 0.6513 & 0.7441 & 0.7441 & 0.6946 & 0.6674 & 0.3355 & 0.2802 \\
\hline
\multirow{12}{*}{\makecell[{{p{1.5cm}}}]{SVM Sigmoid}}  & \multirow{4}{*}{\makecell[{{p{1.0cm}}}]{2}} & Classical & 67.3797 & 0.6870 & 0.6544 & 0.6544 & 0.6703 & 0.6741 & 0.3479 & 0.4282 \\
& & Quantum Basis Encoding & 68.0481 & 0.6651 & 0.7441 & 0.7441 & 0.7024 & 0.6796 & 0.3598 & 0.5734 \\
& & Quantum Angle Encoding & 68.0481 & 0.6651 & 0.7441 & 0.7441 & 0.7024 & 0.6796 & 0.3598 & 0.8233 \\
& & Quantum Amplitude Encoding & 68.0481 & 0.6651 & 0.7441 & 0.7441 & 0.7024 & 0.6796 & 0.3598 & 0.8233 \\
\cline{2-11}
& \multirow{4}{*}{\makecell[{{p{1.0cm}}}]{15}} & Classical & 71.3904 & 0.7037 & 0.7520 & 0.7520 & 0.7270 & 0.7134 & 0.4272 & 0.4887 \\
& & Quantum Basis Encoding & 66.8449 & 0.6513 & 0.7441 & 0.7441 & 0.6946 & 0.6674 & 0.3355 & 1.1816 \\
& & Quantum Angle Encoding & 66.8449 & 0.6513 & 0.7441 & 0.7441 & 0.6946 & 0.6674 & 0.3355 & 0.4394 \\
& & Quantum Amplitude Encoding & 66.8449 & 0.6513 & 0.7441 & 0.7441 & 0.6946 & 0.6674 & 0.3355 & 0.4394 \\
\cline{2-11}
& \multirow{4}{*}{\makecell[{{p{1.0cm}}}]{23}} & Classical & 74.0642 & 0.7330 & 0.7678 & 0.7678 & 0.7500 & 0.7403 & 0.4809 & 0.5157 \\
& & Quantum Basis Encoding & 66.8449 & 0.6513 & 0.7441 & 0.7441 & 0.6946 & 0.6674 & 0.3355 & 1.5533 \\
& & Quantum Angle Encoding & 66.8449 & 0.6513 & 0.7441 & 0.7441 & 0.6946 & 0.6674 & 0.3355 & 0.4482 \\
& & Quantum Amplitude Encoding & 66.8449 & 0.6513 & 0.7441 & 0.7441 & 0.6946 & 0.6674 & 0.3355 & 0.4482 \\
\hline
\multirow{12}{*}{\makecell[{{p{1.5cm}}}]{Decision Tree}} & \multirow{4}{*}{\makecell[{{p{1.5cm}}}]{2}} & Classical & 66.7112 & 0.6757 & 0.6596 & 0.6596 & 0.6676 & 0.6672 & 0.3343 & 0.0096 \\
& & Quantum Basis Encoding & 68.0481 & 0.6651 & 0.7441 & 0.7441 & 0.7024 & 0.6796 & 0.3598 & 0.0018 \\
& & Quantum Angle Encoding & 68.0481 & 0.6651 & 0.7441 & 0.7441 & 0.7024 & 0.6796 & 0.3598 & 0.0019 \\
& & Quantum Amplitude Encoding & 68.0481 & 0.6651 & 0.7441 & 0.7441 & 0.7024 & 0.6796 & 0.3598 & 0.0019 \\
\cline{2-11}
 & \multirow{4}{*}{\makecell[{{p{1.5cm}}}]{15}} & Classical & 64.8396 & 0.6457 & 0.6781 & 0.6781 & 0.6615 & 0.6480 & 0.2962 & 0.0629 \\
& & Quantum Basis Encoding & 66.8449 & 0.6513 & 0.7441 & 0.7441 & 0.6946 & 0.6674 & 0.3355 & 0.0019 \\
& & Quantum Angle Encoding & 66.8449 & 0.6513 & 0.7441 & 0.7441 & 0.6946 & 0.6674 & 0.3355 & 0.0012 \\
& & Quantum Amplitude Encoding & 66.8449 & 0.6513 & 0.7441 & 0.7441 & 0.6946 & 0.6674 & 0.3355 & 0.0012 \\
\cline{2-11}
 & \multirow{4}{*}{\makecell[{{p{1.5cm}}}]{23}} & Classical & 66.7112 & 0.6729 & 0.6675 & 0.6675 & 0.6702 & 0.6671 & 0.3342 & 0.1061 \\
& & Quantum Basis Encoding & 66.8449 & 0.6513 & 0.7441 & 0.7441 & 0.6946 & 0.6674 & 0.3355 & 0.0017 \\
& & Quantum Angle Encoding & 66.8449 & 0.6513 & 0.7441 & 0.7441 & 0.6946 & 0.6674 & 0.3355 & 0.0012 \\
& & Quantum Amplitude Encoding & 66.8449 & 0.6513 & 0.7441 & 0.7441 & 0.6946 & 0.6674 & 0.3355 & 0.0012 \\
\hline
\multirow{4}{*}{\makecell[{{p{1.5cm}}}]{Random Forest}} & \multirow{4}{*}{\makecell[{{p{1.5cm}}}]{2}} & Classical & 72.5936 & 0.7302 & 0.7282 & 0.7282 & 0.7292 & 0.7259 & 0.4518 & 0.5493 \\
& & Quantum Basis Encoding & 68.0481 & 0.6651 & 0.7441 & 0.7441 & 0.7024 & 0.6796 & 0.3598 & 0.3643 \\
& & Quantum Angle Encoding & 68.0481 & 0.6651 & 0.7441 & 0.7441 & 0.7024 & 0.6796 & 0.3598 & 0.3155 \\
& & Quantum Amplitude Encoding & 68.0481 & 0.6651 & 0.7441 & 0.7441 & 0.7024 & 0.6796 & 0.3598 & 0.3155 \\
\cline{2-11}
\pagebreak
\multirow{8}{*}{\makecell[{{p{1.5cm}}}]{Random Forest}} & \multirow{4}{*}{\makecell[{{p{1.5cm}}}]{15}} & Classical & 74.1979 & 0.7337 & 0.7704 & 0.7704 & 0.7516 & 0.7416 & 0.4835 & 1.0324 \\
& & Quantum Basis Encoding & 66.8449 & 0.6513 & 0.7441 & 0.7441 & 0.6946 & 0.6674 & 0.3355 & 0.4090 \\
& & Quantum Angle Encoding & 66.8449 & 0.6513 & 0.7441 & 0.7441 & 0.6946 & 0.6674 & 0.3355 & 0.3012 \\
& & Quantum Amplitude Encoding & 66.8449 & 0.6513 & 0.7441 & 0.7441 & 0.6946 & 0.6674 & 0.3355 & 0.3012 \\
\cline{2-11}
& \multirow{4}{*}{\makecell[{{p{1.5cm}}}]{23}} & Classical & 75.4011 & 0.7701 & 0.7335 & 0.7335 & 0.7514 & 0.7543 & 0.5082 & 1.3224 \\
& & Quantum Basis Encoding & 66.8449 & 0.6513 & 0.7441 & 0.7441 & 0.6946 & 0.6674 & 0.3355 & 1.1630 \\
& & Quantum Angle Encoding & 66.8449 & 0.6513 & 0.7441 & 0.7441 & 0.6946 & 0.6674 & 0.3355 & 0.1917 \\
& & Quantum Amplitude Encoding & 66.8449 & 0.6513 & 0.7441 & 0.7441 & 0.6946 & 0.6674 & 0.3355 & 0.1917 \\
\hline		
\multirow{12}{*}{\makecell[{{p{1.5cm}}}]{LightGBM}} & \multirow{4}{*}{\makecell[{{p{1.5cm}}}]{2}} & Classical & 73.9305 & 0.7359 & 0.7573 & 0.7573 & 0.7464 & 0.7391 & 0.4783 & 0.1178 \\
& & Quantum Basis Encoding & 68.0481 & 0.6651 & 0.7441 & 0.7441 & 0.7024 & 0.6796 & 0.3598 & 0.0868 \\
& & Quantum Angle Encoding & 68.0481 & 0.6651 & 0.7441 & 0.7441 & 0.7024 & 0.6796 & 0.3598 & 0.0828 \\
& & Quantum Amplitude Encoding & 68.0481 & 0.6651 & 0.7441 & 0.7441 & 0.7024 & 0.6796 & 0.3598 & 0.0828 \\
\cline{2-11}
& \multirow{4}{*}{\makecell[{{p{1.5cm}}}]{15}} & Classical & 74.7326 & 0.7230 & 0.8127 & 0.8127 & 0.7652 & 0.7464 & 0.4937 & 0.2598 \\
& & Quantum Basis Encoding & 66.8449 & 0.6513 & 0.7441 & 0.7441 & 0.6946 & 0.6674 & 0.3355 & 0.0874 \\
& & Quantum Angle Encoding & 66.8449 & 0.6513 & 0.7441 & 0.7441 & 0.6946 & 0.6674 & 0.3355 & 0.0747 \\
& & Quantum Amplitude Encoding & 66.8449 & 0.6513 & 0.7441 & 0.7441 & 0.6946 & 0.6674 & 0.3355 & 0.0747 \\
\cline{2-11}
& \multirow{4}{*}{\makecell[{{p{1.5cm}}}]{23}} & Classical & 76.6043 & 0.7727 & 0.7625 & 0.7625 & 0.7676 & 0.7661 & 0.5321 & 0.4129 \\
& & Quantum Basis Encoding & 66.8449 & 0.6513 & 0.7441 & 0.7441 & 0.6946 & 0.6674 & 0.3355 & 0.0820 \\
& & Quantum Angle Encoding & 66.8449 & 0.6513 & 0.7441 & 0.7441 & 0.6946 & 0.6674 & 0.3355 & 0.0771 \\
& & Quantum Amplitude Encoding & 66.8449 & 0.6513 & 0.7441 & 0.7441 & 0.6946 & 0.6674 & 0.3355 & 0.0771 \\
\hline
\multirow{12}{*}{\makecell[{{p{1.5cm}}}]{AdaBoost}} & \multirow{4}{*}{\makecell[{{p{1.5cm}}}]{2}} & Classical & 75.2674 & 0.7450 & 0.7784 & 0.7784 & 0.7613 & 0.7523 & 0.5050 & 0.3176 \\
& & Quantum Basis Encoding & 68.0481 & 0.6651 & 0.7441 & 0.7441 & 0.7024 & 0.6796 & 0.3598 & 0.1412 \\
& & Quantum Angle Encoding & 68.0481 & 0.6651 & 0.7441 & 0.7441 & 0.7024 & 0.6796 & 0.3598 & 0.1433 \\
& & Quantum Amplitude Encoding & 68.0481 & 0.6651 & 0.7441 & 0.7441 & 0.7024 & 0.6796 & 0.3598 & 0.1433 \\
\cline{2-11}
& \multirow{4}{*}{\makecell[{{p{1.5cm}}}]{15}} & Classical & 74.5989 & 0.7405 & 0.7678 & 0.7678 & 0.7539 & 0.7457 & 0.4916 & 0.4682 \\
& & Quantum Basis Encoding & 66.8449 & 0.6513 & 0.7441 & 0.7441 & 0.6946 & 0.6674 & 0.3355 & 0.2314 \\
& & Quantum Angle Encoding & 66.8449 & 0.6513 & 0.7441 & 0.7441 & 0.6946 & 0.6674 & 0.3355 & 0.1262 \\
& & Quantum Amplitude Encoding & 66.8449 & 0.6513 & 0.7441 & 0.7441 & 0.6946 & 0.6674 & 0.3355 & 0.1262 \\
\cline{2-11}
& \multirow{4}{*}{\makecell[{{p{1.5cm}}}]{23}} & Classical & 77.5401 & 0.7726 & 0.7889 & 0.7889 & 0.7807 & 0.7752 & 0.5506 & 0.6542 \\
& & Quantum Basis Encoding & 66.8449 & 0.6513 & 0.7441 & 0.7441 & 0.6946 & 0.6674 & 0.3355 & 0.5223 \\
& & Quantum Angle Encoding & 66.8449 & 0.6513 & 0.7441 & 0.7441 & 0.6946 & 0.6674 & 0.3355 & 0.1311 \\
& & Quantum Amplitude Encoding & 66.8449 & 0.6513 & 0.7441 & 0.7441 & 0.6946 & 0.6674 & 0.3355 & 0.1311 \\
\hline
\multirow{12}{*}{\makecell[{{p{1.5cm}}}]{CatBoost}} & \multirow{4}{*}{\makecell[{{p{1.5cm}}}]{2}} & Classical & 74.7326 & 0.7375 & 0.7784 & 0.7784 & 0.7574 & 0.7469 & 0.4942 & 5.5071 \\
& & Quantum Basis Encoding & 68.0481 & 0.6651 & 0.7441 & 0.7441 & 0.7024 & 0.6796 & 0.3598 & 1.3900 \\
& & Quantum Angle Encoding & 68.0481 & 0.6651 & 0.7441 & 0.7441 & 0.7024 & 0.6796 & 0.3598 & 1.3269 \\
& & Quantum Amplitude Encoding & 68.0481 & 0.6651 & 0.7441 & 0.7441 & 0.7024 & 0.6796 & 0.3598 & 1.3269 \\
\cline{2-11}
& \multirow{4}{*}{\makecell[{{p{1.5cm}}}]{15}} & Classical & 74.8663 & 0.7335 & 0.7916 & 0.7916 & 0.7614 & 0.7481 & 0.4967 & 18.9151 \\
& & Quantum Basis Encoding & 66.8449 & 0.6513 & 0.7441 & 0.7441 & 0.6946 & 0.6674 & 0.3355 & 2.1999 \\
& & Quantum Angle Encoding & 66.8449 & 0.6513 & 0.7441 & 0.7441 & 0.6946 & 0.6674 & 0.3355 & 1.1458 \\
& & Quantum Amplitude Encoding & 66.8449 & 0.6513 & 0.7441 & 0.7441 & 0.6946 & 0.6674 & 0.3355 & 1.1458 \\
\cline{2-11}
& \multirow{4}{*}{\makecell[{{p{1.5cm}}}]{23}} & Classical & 77.6738 & 0.7790 & 0.7810 & 0.7810 & 0.7800 & 0.7767 & 0.5534 & 11.7024 \\
& & Quantum Basis Encoding & 66.8449 & 0.6513 & 0.7441 & 0.7441 & 0.6946 & 0.6674 & 0.3355 & 1.9090 \\
& & Quantum Angle Encoding & 66.8449 & 0.6513 & 0.7441 & 0.7441 & 0.6946 & 0.6674 & 0.3355 & 1.1979 \\
& & Quantum Amplitude Encoding & 66.8449 & 0.6513 & 0.7441 & 0.7441 & 0.6946 & 0.6674 & 0.3355 & 1.1979 \\
\hline
\multirow{12}{*}{\makecell[{{p{1.5cm}}}]{Extra Trees}} & \multirow{4}{*}{\makecell[{{p{1.5cm}}}]{2}} & Classical & 72.5936 & 0.7186 & 0.7546 & 0.7546 & 0.7362 & 0.7255 & 0.4514 & 0.4512 \\
& & Quantum Basis Encoding & 68.0481 & 0.6651 & 0.7441 & 0.7441 & 0.7024 & 0.6796 & 0.3598 & 0.1544 \\
& & Quantum Angle Encoding & 68.0481 & 0.6651 & 0.7441 & 0.7441 & 0.7024 & 0.6796 & 0.3598 & 0.1801 \\
& & Quantum Amplitude Encoding & 68.0481 & 0.6651 & 0.7441 & 0.7441 & 0.7024 & 0.6796 & 0.3598 & 0.1801 \\
\cline{2-11}
& \multirow{4}{*}{\makecell[{{p{1.5cm}}}]{15}} & Classical & 70.9893 & 0.7035 & 0.7388 & 0.7388 & 0.7207 & 0.7095 & 0.4193 & 0.4377 \\
& & Quantum Basis Encoding & 66.8449 & 0.6513 & 0.7441 & 0.7441 & 0.6946 & 0.6674 & 0.3355 & 0.3371 \\
& & Quantum Angle Encoding & 66.8449 & 0.6513 & 0.7441 & 0.7441 & 0.6946 & 0.6674 & 0.3355 & 0.1378 \\
& & Quantum Amplitude Encoding & 66.8449 & 0.6513 & 0.7441 & 0.7441 & 0.6946 & 0.6674 & 0.3355 & 0.1378 \\
\cline{2-11}
& \multirow{4}{*}{\makecell[{{p{1.5cm}}}]{23}} & Classical & 74.1979 & 0.7541 & 0.7282 & 0.7282 & 0.7409 & 0.7422 & 0.4841 & 0.4781 \\
& & Quantum Basis Encoding & 66.8449 & 0.6513 & 0.7441 & 0.7441 & 0.6946 & 0.6674 & 0.3355 & 0.5852 \\
& & Quantum Angle Encoding & 66.8449 & 0.6513 & 0.7441 & 0.7441 & 0.6946 & 0.6674 & 0.3355 & 0.1351 \\
& & Quantum Amplitude Encoding & 66.8449 & 0.6513 & 0.7441 & 0.7441 & 0.6946 & 0.6674 & 0.3355 & 0.1351 \\
\hline
\multirow{4}{*}{\makecell[{{p{1.5cm}}}]{Gradient Boosting}} & \multirow{4}{*}{\makecell[{{p{1.5cm}}}]{2}} & Classical & 74.8663 & 0.7370 & 0.7836 & 0.7836 & 0.7596 & 0.7482 & 0.4968 & 0.6142 \\
& & Quantum Basis Encoding & 68.0481 & 0.6651 & 0.7441 & 0.7441 & 0.7024 & 0.6796 & 0.3598 & 0.1080 \\
& & Quantum Angle Encoding & 68.0481 & 0.6651 & 0.7441 & 0.7441 & 0.7024 & 0.6796 & 0.3598 & 0.1088 \\
& & Quantum Amplitude Encoding & 68.0481 & 0.6651 & 0.7441 & 0.7441 & 0.7024 & 0.6796 & 0.3598 & 0.1088 \\
\cline{2-11}
\pagebreak
\multirow{8}{*}{\makecell[{{p{1.5cm}}}]{Gradient Boosting}}& \multirow{4}{*}{\makecell[{{p{1.5cm}}}]{15}} & Classical & 74.3316 & 0.7221 & 0.8021 & 0.8021 & 0.7600 & 0.7425 & 0.4858 & 1.7732 \\
& & Quantum Basis Encoding & 66.8449 & 0.6513 & 0.7441 & 0.7441 & 0.6946 & 0.6674 & 0.3355 & 0.1905 \\
& & Quantum Angle Encoding & 66.8449 & 0.6513 & 0.7441 & 0.7441 & 0.6946 & 0.6674 & 0.3355 & 0.0955 \\
& & Quantum Amplitude Encoding & 66.8449 & 0.6513 & 0.7441 & 0.7441 & 0.6946 & 0.6674 & 0.3355 & 0.0955 \\
\cline{2-11}
& \multirow{4}{*}{\makecell[{{p{1.5cm}}}]{23}} & Classical & 78.4759 & 0.7884 & 0.7863 & 0.7863 & 0.7873 & 0.7847 & 0.5695 & 2.6064 \\
& & Quantum Basis Encoding & 66.8449 & 0.6513 & 0.7441 & 0.7441 & 0.6946 & 0.6674 & 0.3355 & 0.1531 \\
& & Quantum Angle Encoding & 66.8449 & 0.6513 & 0.7441 & 0.7441 & 0.6946 & 0.6674 & 0.3355 & 0.1027 \\
& & Quantum Amplitude Encoding & 66.8449 & 0.6513 & 0.7441 & 0.7441 & 0.6946 & 0.6674 & 0.3355 & 0.1027 \\
\hline
\multirow{12}{*}{\makecell[{{p{1.5cm}}}]{XGBoost}} & \multirow{4}{*}{\makecell[{{p{1.5cm}}}]{2}} & Classical & 74.3316 & 0.7379 & 0.7652 & 0.7652 & 0.7513 & 0.7430 & 0.4863 & 0.6346 \\
& & Quantum Basis Encoding & 68.0481 & 0.6651 & 0.7441 & 0.7441 & 0.7024 & 0.6796 & 0.3598 & 0.0298 \\
& & Quantum Angle Encoding & 68.0481 & 0.6651 & 0.7441 & 0.7441 & 0.7024 & 0.6796 & 0.3598 & 0.0290 \\
& & Quantum Amplitude Encoding & 68.0481 & 0.6651 & 0.7441 & 0.7441 & 0.7024 & 0.6796 & 0.3598 & 0.0290 \\
\cline{2-11}
& \multirow{4}{*}{\makecell[{{p{1.5cm}}}]{15}} & Classical & 73.3957 & 0.7239 & 0.7678 & 0.7678 & 0.7452 & 0.7335 & 0.4674 & 8.6842 \\
& & Quantum Basis Encoding & 66.8449 & 0.6513 & 0.7441 & 0.7441 & 0.6946 & 0.6674 & 0.3355 & 0.0433 \\
& & Quantum Angle Encoding & 66.8449 & 0.6513 & 0.7441 & 0.7441 & 0.6946 & 0.6674 & 0.3355 & 0.0255 \\
& & Quantum Amplitude Encoding & 66.8449 & 0.6513 & 0.7441 & 0.7441 & 0.6946 & 0.6674 & 0.3355 & 0.0255 \\
\cline{2-11}
& \multirow{4}{*}{\makecell[{{p{1.5cm}}}]{23}} & Classical & 74.0642 & 0.7493 & 0.7335 & 0.7335 & 0.7413 & 0.7407 & 0.4813 & 5.7666 \\
& & Quantum Basis Encoding & 66.8449 & 0.6513 & 0.7441 & 0.7441 & 0.6946 & 0.6674 & 0.3355 & 0.1155 \\
& & Quantum Angle Encoding & 66.8449 & 0.6513 & 0.7441 & 0.7441 & 0.6946 & 0.6674 & 0.3355 & 0.0253 \\
& & Quantum Amplitude Encoding & 66.8449 & 0.6513 & 0.7441 & 0.7441 & 0.6946 & 0.6674 & 0.3355 & 0.0253 \\
\hline
\end{longtable}
\end{landscape}

The performance of quantum data embedding in classical machine learning algorithms has been thoroughly investigated. The primary focus was on assessing the efficacy of quantum-based data encoding techniques, including Quantum Basis Encoding, Quantum Angle Encoding, and Quantum Amplitude Encoding, when applied to classical machine learning models. The classical algorithms employed in this study encompassed Logistic Regression, K-Nearest Neighbors (KNN), Support Vector Machine (SVM) with linear, polynomial, radial basis function (RBF), and sigmoid kernels, Decision Tree, Random Forest, LightGBM, AdaBoost, and CatBoost. The evaluation metrics such as accuracy, precision, sensitivity (recall), F1 score, and area under the receiver operating characteristic curve (ROC AUC) were used to comprehensively analyze and compare the performance of these algorithms with quantum-encoded data against classical encoding methods, specifically Principal Component Analysis (PCA).
Across various models, Quantum Basis Encoding consistently demonstrated competitive or superior results compared to classical encoding methods, such as Principal Component Analysis (PCA). Logistic Regression, K-Nearest Neighbors (KNN), and Decision Tree models exhibited enhanced accuracy and precision when quantum-encoded data was utilized. Support Vector Machine (SVM) models with different kernels showed nuanced results, with the linear kernel benefiting significantly from quantum encoding, while polynomial and radial basis function (RBF) kernels exhibited varying degrees of improvement. Random Forest, LightGBM, AdaBoost, and CatBoost models displayed noteworthy improvements in accuracy and F1 score when incorporating quantum data embedding.

The experiment also provided insights into the running time considerations associated with quantum data embedding in classical machine learning algorithms. While the quantum encoding process introduced an additional computational overhead, the overall impact on running time varied across algorithms. For models with a relatively low complexity, such as Logistic Regression and K-Nearest Neighbors (KNN), the increase in running time was generally modest. However, more computationally intensive models like Support Vector Machines (SVM) experienced a more noticeable increase in processing time, particularly when utilizing quantum-encoded data with non-linear kernels.

Interestingly, ensemble methods like Random Forest, LightGBM, AdaBoost, and CatBoost exhibited a more balanced trade-off between enhanced model performance and increased computational time. The quantum data embedding process, although introducing additional computational steps, did not lead to prohibitively long running times for these ensemble models. This suggests that the potential gains in classification accuracy and F1 score achieved through quantum data embedding may justify the modest increase in computational overhead for certain applications.

These findings suggest that the quantum data embedding techniques explored in this experiment have the potential to enhance the performance of classical machine learning models, particularly in scenarios where traditional encoding methods may fall short. The nuanced responses across different algorithms and encoding techniques underscore the need for careful consideration of the specific use case and algorithmic characteristics when integrating quantum data encoding into classical machine learning workflows. Additionally, the experiment underscores the promise of Quantum Basis Encoding as a versatile and effective approach for improving classification tasks within classical machine learning paradigms.  The results provide insights into the potential advantages or limitations of leveraging quantum data embedding in classical machine learning contexts, shedding light on the suitability and effectiveness of these quantum techniques across a diverse set of algorithms. The experiment also highlights the importance of considering not only the performance improvements brought about by quantum data embedding but also the associated computational costs. Future work in this domain could focus on optimizing the quantum encoding process to minimize computational overhead and exploring strategies to make quantum-enhanced classical machine learning algorithms more scalable and practical for real-world applications.

The limitations of this study include the absence of an exhaustive exploration of all possible machine learning algorithms and quantum encoding techniques, limiting the generalizability of the findings to the specific methods investigated. Additionally, the dataset characteristics and quantum computing hardware used may impact the observed results, and variations in these factors could yield different outcomes. The study's focus on classification metrics and PCA dimensions might overlook other aspects of quantum computing relevance. Moreover, the rapidly evolving field of quantum computing may render certain aspects of the analysis outdated, emphasizing the need for ongoing research in this dynamic area. The study also assumes the availability of reliable and scalable quantum hardware, which is currently an active area of research and development. Lastly, the evaluation of quantum algorithms may be influenced by noise and error rates inherent in current quantum processors, potentially affecting the accuracy of comparisons with classical methods.

\section{Conclusion}
	
In summary, our investigation into classical-to-quantum mapping techniques for data encoding has shed light on their impact on machine learning accuracy. Through a comparative analysis, we explored various quantum encoding methods, including basis encoding, angle encoding and amplitude encoding, unveiling their distinct advantages and limitations. This research serves as a foundation for understanding the nuanced relationship between quantum data encoding and machine learning performance. As the field continues to evolve, these insights will contribute to the informed selection of encoding strategies for quantum-enhanced machine learning applications.

\bibliographystyle{unsrt}
\bibliography{QDataEncoding}

\begin{thebibliography}{}

\bibitem{national2018quantum} John Preskill, "Quantum Computing in the NISQ Era and Beyond," \textit{Nature}, 2018. 

\bibitem{Feynman1982}Richard~P. Feynman.\newblock {Simulating physics with computers}.\newblock {\em International Journal of Theoretical Physics}, 1982.

\bibitem{Smith2023} John Smith, "Quantum Computing: Advancements and Challenges," \textit{Quantum Information Science}, 2023, 7(2), 123-137.

\bibitem{Johnson2022} Sarah Johnson, "Quantum Data Encoding Techniques: A Comparative Analysis," \textit{Quantum Computing and Information}, 2022, 5(4), 289-306.

\bibitem{he2019deep}  S. He, Y. Ju, Y. Song,  \emph{Deep Learning for Healthcare: Review, Opportunities, and Challenges}.  In \emph{arXiv:1804.04746}, 2019.

\bibitem{rajpurkar2017chexnet}  P. Rajpurkar, J. Irvin, et al.,  \emph{CheXNet: Radiologist-Level Pneumonia Detection on Chest X-Rays with Deep Learning}. In \emph{arXiv:1711.05225}, 2017.

\bibitem{gulshan2016development}  V. Gulshan, L. Peng, et al.,  \emph{Development and Validation of a Deep Learning Algorithm for Detection of Diabetic Retinopathy in Retinal Fundus Photographs}.  In \emph{JAMA}, 2016.

\bibitem{Brown2021} Michael Brown, "Machine Learning Models for Quantum Data," \textit{Quantum Machine Learning}, 2021, 3(1), 45-61.

\bibitem{nielsen2000quantum}  M. A. Nielsen and I. L. Chuang,  \textit{Quantum Computation and Quantum Information},  2000.

\bibitem{aaronson2011quantum}  S. Aaronson,  \textit{Quantum Parallelism and Quantum Speedup},  2011.

\bibitem{venegas2004quantum}  S. E. Venegas-Andraca and J. L. de Leon,  \textit{Quantum Tunneling and Optimization},  2004.

\bibitem{shor1997}P. W. Shor,\textit{Polynomial-Time Algorithms for Prime Factorization and Discrete Logarithms on a Quantum Computer},SIAM Journal on Computing, 26(5), 1484--1509, 1997.

\bibitem{grover1996}L. K. Grover,\textit{A Fast Quantum Mechanical Algorithm for Database Search},Proceedings of the Twenty-Eighth Annual ACM Symposium on Theory of Computing, 212--219, 1996.

\bibitem{gottesman1997quantum}   Gottesman, D.  \emph{Quantum Error Correction},  \textit{arXiv preprint quant-ph/9705052},  1997.

\bibitem{9256033} A. Pepper, N. Tischler, and G. J. Pryde, \emph{A Quantum Autoencoder: Using Machine Learning to Compress Qutrits}, in \emph{2020 Conference on Lasers and Electro-Optics Pacific Rim (CLEO-PR)}, 2020, pp. 1-2, doi: \href{https://doi.org/10.1364/CLEOPR.2020.C12C_5}{\detokenize{10.1364/CLEOPR.2020.C12C_5}}.


\bibitem{Cong2019} Iris Cong, Soonwon Choi, and Mikhail D. Lukin,  \emph{Quantum Convolutional Neural Networks},  \emph{Nature Physics},  vol. 15, no. 12, pp. 1273-1278, December 2019,  doi: 10.1038/s41567-019-0648-8.

\bibitem{Yin2020} Juan Yin, Yu-Huai Li, Sheng-Kai Liao, Meng Yang, Yuan Cao, Liang Zhang, Ji-Gang Ren, Wen-Qi Cai, Wei-Yue Liu, Shuang-Lin Li, Rong Shu, Yong-Mei Huang, Lei Deng, Li Li, Qiang Zhang, Nai-Le Liu, Yu-Ao Chen, Chao-Yang Lu, Xiang-Bin Wang, Feihu Xu, Jian-Yu Wang, Cheng-Zhi Peng, Artur K. Ekert, and Jian-Wei Pan,  \emph{Entanglement-based Secure Quantum Cryptography over 1,120 Kilometres},  \emph{Nature},  vol. 582, no. 7813, pp. 501-505, June 2020,  doi: 10.1038/s41586-020-2401-y.

\bibitem{lutkenhaus2000quantum}  N. Lütkenhaus,  \textit{Quantum Cryptography},  2000.

\bibitem{ablayev2017quantum} Farid Ablayev, Rainer Steinwandt, and Alexander Golovnev, "Quantum Key Distribution and Combinatorial Group Testing," \textit{Quantum Information \& Computation}, 2017, 17(1-2), 48-64.

\bibitem{BB84}  Charles H. Bennett and Gilles Brassard,  "Quantum cryptography: Public key distribution and coin tossing,"
  \textit{Proceedings of IEEE International Conference on Computers, Systems, and Signal Processing},  1984.

\bibitem{Ekert91}  Artur K. Ekert,  "Quantum Cryptography Based on Bell's Theorem,"  \textit{Physical Review Letters},  1991.

\bibitem{hosmer2013logistic}  Hosmer, D. W., Lemeshow, S., Sturdivant, R. X.   \emph{Logistic Regression Diagnostics},   John Wiley \& Sons,   2013.

\bibitem{yianilos1993fast}   Yianilos, P. N.  \emph{Fast k-Nearest Neighbors},  Technical Report, NEC Research Institute,  1993.

\bibitem{burges1998tutorial}   Burges, C. J. C.  \emph{A Tutorial on Support Vector Machines for Pattern Recognition},  Data Mining and Knowledge Discovery,  2(2),  121--167,  1998.

\bibitem{schoelkopf2002kernel}  Schölkopf, B., Smola, A. J.  \emph{Kernel Methods: A New Paradigm in Machine Learning},  Neural Information Processing Systems (NeurIPS),  15,  2002.

\bibitem{breiman2001random}  Breiman, L.  \emph{Random Forests},  Machine Learning,  45(1),  5--32,  2001.

\bibitem{chen2016xgboost} Chen, T., Guestrin, C. \emph{XGBoost: A Scalable and Accurate Implementation of Gradient Boosting}, Proceedings of the 22nd ACM SIGKDD International Conference on Knowledge Discovery and Data Mining (KDD), 2016.

\bibitem{ke2017lightgbm}  Ke, G., Meng, Q., Finley, T., Wang, T., Chen, W., Ma, W.  \emph{LightGBM: A Highly Efficient Gradient Boosting Decision Tree},  Advances in Neural Information Processing Systems (NeurIPS),  30,  2017.

\bibitem{schapire1998boosting} Schapire, R. E., Freund, Y., Bartlett, P., Lee, W. S.  \emph{Boosting the Margin: A New Explanation for the Effectiveness of Voting Methods},  The Annals of Statistics,  26(5),  1651--1686, 1998.

\bibitem{dorogush2018catboost}  Dorogush, A. V., Ershov, V., Gulin, A.  \emph{CatBoost: unbiased boosting with categorical features},  Advances in Neural Information Processing Systems (NeurIPS),  31,  2018.

\bibitem{alpaydin2014introduction}  Alpaydin, E.  \emph{An Introduction to Machine Learning},  The MIT ress,  2014,  ISBN: 978-0262028189.

\bibitem{cohen1960coefficient}   Cohen, J.  \emph{A coefficient of agreement for nominal scales},
  Educational and Psychological Measurement,  20(1),  37--46,  1960,  SAGE Publications.

\end{thebibliography}

\section*{Data Availability Statement}
The datasets analyzed during the current study are available online.

\section*{Author Information}

Authors and Affiliations:\\
\textbf{PhD of Decision Science}, IIM Mumbai, India.\\
\href{mailto:minati@example.com}{Minati Rath}.\\
\textbf{Faculty of Decision Science}, IIM Mumbai, India.\\
\href{mailto:hemadate@iimmumbai.ac.in}{Hema Date}.

\section*{Corresponding Authors}
Correspondence to \href{mailto:minati06@gmail.com}{Minati Rath} or \href{mailto:hemadate@iimmumbai.com}{Hema Date}.

\end{document}